\documentstyle[12pt]{article}
\newlength{\cmd}
\topmargin - 10 mm

\begin{document}

\vspace{20  mm}

\begin{center}
{\large {\bf Traveling waves in reaction-diffusion system: }} \vspace{3 mm }

{\large {\bf diffusion with finite velocity and}} \vspace{3  mm }

{\large {\bf Kolmogorov-Petrovskii-Piskunov kinetics }} \vspace{3  mm }

{\large {\bf {\vspace{10  mm} Sergei Fedotov \vspace{5  mm } }}}

{\large {\bf PIK, Telegrafenberg C4 \\}}

{\large {\bf Potsdam 14412 Deutschland}}

\vspace{3  mm }

{\large {\bf E-mail: fedotov@pik-potsdam.de\\}}

\end{center}

\vspace{25  mm}

\begin{center}
{\large {\bf {ABSTRACT}}}
\end{center}

\vspace{10  mm}

\baselineskip \cmd

A new asymptotic method is presented for the analysis of the
traveling waves in the one-dimensional reaction-diffusion system with the
diffusion with a finite velocity and Kolmogorov-Petrovskii-Piskunov
kinetics. The analysis makes use of the path-integral approach, scaling procedure and  the singular
perturbation techniques involving the large deviations theory for the
Poisson random walk. The exact formula for the position and speed of  
reaction front is derived. It is found that the reaction front dynamics is formally associated with the relativistic Hamiltonian/Lagrangian mechanics.


\newpage

Phenomena of the wave propagation in the non-equilibrium media described by
reaction-diffusion equations have attracted considerable interest in a wide
variety of scientific fields including physics, chemistry, biology, etc. The
excellent reviews of the work in this area can be found in the books [1-4].
Of fundamental interest is the rate at which the wave propagates through the
non-linear dissipative system. The basic common feature in many examples is
that the transport process determining the propagation rate is described by
the conventional diffusion (Fick's law). In this case the propagation
velocity $u$ can be found from a simple dimensional analysis, that is $u\sim 
\sqrt{DU},$ where $D$ is the diffusion coefficient and $U$ is the
characteristic reaction rate constant.

It is well known that the diffusion approximation gives rise to the infinite
speed of heat/mass propagation that is if a sudden change of
temperature/concentration takes place somewhere in the space, it will be
felt immediately everywhere with the exponentially small amplitude. It is
therefore desirable to have a theory for the nonlinear wave propagation in
which the boundness of the transport process would be taken into account.
The purpose of this Letter is to present such a theory giving a new
asymptotic method for calculating the propagation speed for the traveling
wave in the reaction-diffusion system involving the diffusion with a finite
velocity [5-10] and the chemical kinetics of the
Kolmogorov-Petrovskii-Piskunov (KPP) type [1-4, 11-17].

Our starting point is a phenomenological system of the one-dimensional equations for the time
evolution of the scalar field $\rho $ and its flux $J$ 
$$
\frac{\partial \rho }{\partial t}+\frac{\partial J}{\partial x}=U\rho \left(
1-\rho \right) ,\eqno(1) 
$$

$$
\frac{\partial J}{\partial t}=-\frac{J-J_0}\tau ,\hspace{.3in}J_0=-D\frac{%
\partial \rho }{\partial x},\eqno(2) 
$$
where $U$ is the reaction rate constant corresponding to the KPP-kinetics, $%
D $ is the diffusion coefficient corresponding to the Fick's law and $\tau $
is the relaxation time. When $U=0,$ the system (1),(2) reduces to the
telegraph equation [5-10] 
$$
\tau \frac{\partial ^2\rho }{\partial t^2}+\frac{\partial \rho }{\partial t}%
=D\frac{\partial ^2\rho }{\partial x^2},\eqno(3) 
$$
when $\tau =0,$ we have the classical KPP-equation [1-4] 
$$
\frac{\partial \rho }{\partial t}=D\frac{\partial ^2\rho }{\partial x^2}%
+U\rho \left( 1-\rho \right) .\eqno(4) 
$$

If we now solve the equation (2) with the initial condition $J\left(
0,x\right) =0$ we may eliminate $J$ from (1) to obtain the single equation
for $\rho $%
$$
\frac{\partial \rho }{\partial t}=\frac D\tau \int_0^t\exp \left( -\frac{t-s}%
\tau \right) \frac{\partial ^2\rho \left( s,x\right) }{\partial x^2}ds+U\rho
\left( 1-\rho \right) .\eqno(5) 
$$
This equation may be considered as a generalization of the KPP-equation (4)
to the case in which the finite speed of the transport process is taken into
account ($\tau \neq 0$).

We specify the following initial condition 
$$
\rho \left( 0,x\right) =\theta \left( x\right) ,\eqno(6) 
$$
where $\theta \left( x\right) $ is a Heaviside function $\theta \left(
x\right) =1$ for $x<0$ and $\theta \left( x\right) =0$ for $x>0.$

The basic problem is to find the traveling wave solution $\psi \left(
x-ut\right) $ to the problem (5), (6), where $\psi \left( z\right) $ is a
monotonically decreasing function such that $\psi \left( -\infty \right) =1$
and $\psi \left( \infty \right) =0,$ and $u$ is the speed at which the wave
profile $\psi $ moves in the positive $x$-direction. For the KPP-equation
(4) ($\tau =0$) with the initial condition (6) the traveling wave moves
with the velocity $u=\sqrt{4DU}$ [1-4]. We expect that for (5), (6) the
speed $u=\sqrt{4DU}f\left( \tau U\right), $ where $f\left( z\right) $  is the dimensionless function such that $f\left( 0\right) =1$.
It should be noted that our method of calculation will be
nonperturbative in a sense that we do not treat the relaxation time $\tau $
as a small parameter.

We are interested in the long-time large-distance behavior of the traveling
wave solution of (5), (6) as $t\rightarrow \infty $ and $x\rightarrow \infty 
$ . It is convenient therefore to make the scaling [12-17] 
$$
t\rightarrow \frac t\varepsilon ,\hspace{.5in}x\rightarrow \frac
x\varepsilon \eqno(7) 
$$
and rewrite the Cauchy problem (5),(6) for $\rho ^\varepsilon \left(
t,x\right) =\rho \left( \frac t\varepsilon ,\frac x\varepsilon \right) $ in
the following form 
$$
\frac{\partial \rho ^\varepsilon }{\partial t}=\frac D\tau \int_0^t\exp
\left( -\frac{t-s}{\varepsilon \tau }\right) \frac{\partial ^2\rho
^\varepsilon \left( s,x\right) }{\partial x^2}ds+\frac U\varepsilon \rho
^\varepsilon \left( 1-\rho ^\varepsilon \right) ,\hspace{.3in}\rho
^\varepsilon \left( 0,x\right) =\theta \left( x\right) .\eqno(8) 
$$

It is clear from this equation that the scaling (7) describing a
simultaneous contraction of time and space alternatively corresponds to the
rapid chemical reaction and slow transport process. We expect that after
rescaling the wave profile develops into the reaction front: $\rho
^\varepsilon \left( t,x\right) =\psi \left( \frac{x-ut}\varepsilon \right) $
tends to a unit step function $\theta \left( x-ut\right) $ as $\varepsilon
\rightarrow 0$.

Our goal is now to find a function $G\left( t,x\right) $ determining the
position of the reaction front that is 
$$
\ \lim_{\epsilon \to 0}\,\,\rho ^\varepsilon \left( t,x\right) =
\left\{\begin{array}{ll}
0 & \mbox{if $ G\left( t,x\right) < 0 $}\\
1 & \mbox{otherwise.}
\end{array}
\right.
\eqno(9) 
$$

In this Letter we restrict ourselves in finding the upper bound for $\rho
^\varepsilon \left( t,x\right) $ in the form 
$$
\rho ^\varepsilon \left( t,x\right) \leq \exp \left\{ \frac{G\left(
t,x\right) }\varepsilon \right\} \hspace{.2in}as\hspace{.2in}\varepsilon
\rightarrow 0.\eqno(10) 
$$
It is clear from (10) that $\rho ^\varepsilon \left( t,x\right) \rightarrow
0 $ if $G\left( t,x\right) <0$ and $\varepsilon \rightarrow 0.$ One can
prove that $\rho ^\varepsilon \left( t,x\right) \rightarrow 1 $ if $G\left(
t,x\right) >0$ and $\varepsilon \rightarrow 0.$

It follows from the property of the KPP-kinetics in (8) that 
$$
\rho ^\varepsilon \left( t,x\right) \leq \varphi ^\varepsilon \left(
t,x\right) \exp \left( \frac{Ut}\varepsilon \right) ,\eqno(11) 
$$
where $\varphi ^\varepsilon \left( t,x\right) $ is a solution of the linear
initial problem 
$$
\frac{\partial \varphi ^\varepsilon }{\partial t}=\frac D\tau \int_0^t\exp
\left( -\frac{\left( U+\tau ^{-1}\right) \left( t-s\right) }\varepsilon
\right) \frac{\partial ^2\varphi ^\varepsilon \left( s,x\right) }{\partial
x^2}ds,\hspace{.3in}\varphi ^\varepsilon \left( 0,x\right) =\theta \left(
x\right) .\eqno(12) 
$$

Our strategy to find the function $G\left( t,x\right) $ is to analyze the
above Cauchy problem in terms of the probability theory and thereby to obtain an
estimate of $\varphi ^\varepsilon \left( t,x\right) $ in the limit $%
\varepsilon \rightarrow 0$. The basic idea is that we can deal with the
problem (12) in terms of the random walks of Poisson type [7-9]. If we
introduce the notations $c$ (velocity) and $\nu $ (frequency) such that 
$$
c^2=\frac D\tau ,\hspace{.3in}2\nu =U+\frac 1\tau ,\eqno(13)
$$
than the solution of linear initial value problem (12) can be written as an
expectation value of the initial distribution $\theta $ [7,8] 
$$
\varphi ^\varepsilon \left( t,x\right) ={\bf E}\theta \left( x\left(
t\right) \right) ,\eqno(14)
$$
where ${\bf E}$ is the expectation operator and $x\left( t\right) $ is a
random Poisson walk, i.e. a solution of the stochastic differential equation 
$$
\frac{dx}{ds}=v\left( \frac s\varepsilon \right) ,\hspace{.3in}x\left(
0\right) =x,\hspace{.3in}0<s<t,\eqno(15)
$$
where $v\left( s\right) $ is the Markovian dichotomous velocity taking only
two values $\left\{ c,-c\right\} $ with the frequency $\nu $ [18]. From the
probabilistic point of view the key factor underlying the nonlocal character
of (12) is that the dynamics of $x\left( t\right) $ is non-Markovian [18].
To obtain an estimate of $\varphi ^\varepsilon \left( t,x\right) $ as $%
\varepsilon \rightarrow 0$ we need to know an explicit expression for $%
\varphi ^\varepsilon \left( t,x\right) $ as a path-integral [19] 
$$
\varphi ^\varepsilon \left( t,x\right) =\int \theta \left( x\left( t\right)
\right) P\left[ x\left( \cdot \right) \right] {\cal D}x,\eqno(16)
$$
where $P\left[ x\left( \cdot \right) \right] $ is a conditional probability
density functional for the random process $x\left( s\right) $ 
$$
P\left[ x\left( \cdot \right) \right] =\int \delta \left[ \frac{dx}{ds}%
-v\left( \frac s\varepsilon \right) \right] P\left[ v\left( \cdot \right)
\right] {\cal D}v,
$$
where $\delta \left[ \cdot \right] $ is the $\delta -$functional that is the
extension of the ordinary $\delta -$function to the functional integration
[19]. By using (16) and the formalism based on the auxiliary function $u$
[20,21] we can write down the following expression for $\varphi ^\varepsilon
\left( t,x\right) $ 
$$
\varphi ^\varepsilon \left( t,x\right) =\int \int \int \theta \left( x\left(
t\right) \right) \exp \left( i\int_0^tu\left( s\right) \left( \frac{dx}{ds}%
-v\left( \frac s\varepsilon \right) \right) ds\right) P\left[ v\left( \cdot
\right) \right] {\cal D}v{\cal D}u{\cal D}x.\eqno(17)
$$

In the ''weak noise limit'' $\varepsilon \rightarrow 0$ one can get the
following estimate for $\varphi ^\varepsilon \left( t,x\right) $ (the
details of calculation involving the large deviations theory for the Poisson
random walk will appear elsewhere) 
$$
\varphi ^\varepsilon \left( t,x\right) \sim \exp \left\{ -\frac 1\varepsilon
\min_{x\left( 0\right) =x;x\left( t\right) =0}\int_0^t\left( p\frac{dx}{ds}%
-H\left( p\right) \right) ds\right\} ,\eqno(18) 
$$
where the function $H\left( p\right) $ has the form of the relativistic
Hamiltonian [22] 
$$
H\left( p\right) =c\sqrt{m^2c^2+p^2}-\nu \eqno(19) 
$$
with the ''effective mass'' $m=\nu c^{-2}.$ One can rewrite (18) in terms of
the relativistic Lagrangian [22] 
$$
\varphi ^\varepsilon \left( t,x\right) \sim \exp \left\{ -\frac 1\varepsilon
\min_{x\left( 0\right) =x;x\left( t\right) =0}\int_0^tLds\right\} ,\eqno(20) 
$$
where 
$$
L=-mc^2\sqrt{1-\frac 1{c^2}\left( \frac{dx}{ds}\right) ^2}+\nu .\eqno(21) 
$$

We are now in a position to complete the derivation of the function $G\left(
t,x\right) $ determining the reaction front position and its speed. One finds
after straightforward calculation that the optimal trajectory giving the
minimum in (20) is $x\left( s\right) =-\frac xts+x$ and the corresponding
minimal action is $mc^2t\sqrt{1-\frac 1{c^2}\left( \frac xt\right) ^2}$. By
using (10), (11) and (20) and the relation $m=\nu c^{-2}$ we obtain 
$$
G\left( t,x\right) =Ut-\nu t+\nu t\sqrt{1-\frac 1{c^2}\left( \frac xt\right)
^2}.\eqno(22) 
$$

Equating $G\left( t,x\right) $ to $0$ we obtain the position of reaction
front $x\left( t\right) $%
$$
x\left( t\right) =ut,\hspace{.3in}u=c\sqrt{1-\left( \frac{\nu -U}\nu \right)
^2}, \hspace{.3in}U\leq \nu .\eqno(23)
$$
Taking into account (13) the speed of the reaction front $u$ can be
rewritten in terms of the phenomenological parameters $D$ and $\tau $ 
$$
u=\sqrt{\frac{4DU}{\left( 1+\tau U\right) ^2}},\hspace{.3in}\tau U\leq 1.%
\eqno(24)
$$
It follows from (23) and (24) that the speed $u$ takes the maximum value $%
c=\sqrt{\frac D\tau },$ when $\tau U=1$ or $U=\nu $. If the velocity of
propagation $c$ were infinitely great and the time $\tau $ were infinitely
small such that $D=c^2\tau =const$ (the diffusion approximation for the
random walk of Poisson type), equations (23) and (24) would merely give $u=%
\sqrt{4DU}$ - the classical result of the KPP-theory [1-4]. 

In summury, we have derived the {\it exact} formula for the position and speed of the
reaction front in the one-dimensional dissipative system involving the
diffusion with a finite velocity and the KPP-kinetics.  It has been found
that the reaction front dynamics for such a system can be formally
associated with the relativistic Hamiltonian/Lagrangian mechanics. There are
several possible directions to explore by the method developed here. First
one may study the influence of non-uniform distribution of the reaction rate
constant $U$ which might induce the jumps of reaction fronts [11]. One can
also extend the analysis to describe the interaction between the turbulent
diffusion with a finite velocity [10] and the KPP-kinetics in the
three-dimensional space [17].

The author would like to thank Rupert Klein and Alexander S. Mikhailov for
interesting and helpful discussions. The research was supported in part by
DFG Project KL 611/5-1.

\end{document}